\documentclass{emulateapj}
\usepackage{natbib}
\usepackage{epsfig}
\usepackage{graphicx}
\usepackage{subfigure}
\usepackage{float}
\usepackage{amsmath}
\usepackage{color}
\usepackage{amssymb}
\usepackage{amsfonts}
\usepackage{apjfonts}
\usepackage[colorlinks,linkcolor=blue,anchorcolor=green,citecolor=blue]{hyperref}
\shorttitle{Constraining Anisotropic Lorentz Violation via GRB 160625B}
\shortauthors{Wei et al.}
\begin{document}

\title{Constraining Anisotropic Lorentz Violation
via the Spectral-Lag Transition of GRB 160625B}

\author{Jun-Jie Wei,\altaffilmark{1,2}
Xue-Feng Wu,\altaffilmark{1,3,4}
Bin-Bin Zhang,\altaffilmark{5,6}
Lang Shao,\altaffilmark{7,1}
Peter M{\'e}sz{\'a}ros,\altaffilmark{8,9,10}
and V.\ Alan Kosteleck{\'y}\altaffilmark{11}}

\affil{$^1$Purple Mountain Observatory, Chinese Academy of Sciences,
Nanjing 210008, China; xfwu@pmo.ac.cn \\
$^2$ Guangxi Key Laboratory for Relativistic Astrophysics,
Nanning 530004, China \\
$^3$School of Astronomy and Space Science,
University of Science and Technology of China, Hefei, Anhui 230026, China\\
$^4$Joint Center for Particle, Nuclear Physics and Cosmology,
Nanjing University-Purple Mountain Observatory, Nanjing 210008, China\\
$^5$Instituto de Astrof\'isica de Andaluc\'a (IAA-CSIC),
P.O. Box 03004, E-18080 Granada, Spain \\
$^6$Scientist Support LLC, Madsion, AL 35758, USA \\
$^7$Department of Space Sciences and Astronomy, Hebei Normal University,
Shijiazhuang 050024, China\\
$^8$Department of Astronomy and Astrophysics, Pennsylvania State University,
525 Davey Laboratory, University Park, PA 16802, USA\\
$^9$Department of Physics, Pennsylvania State University,
104 Davey Laboratory, University Park, PA 16802, USA\\
$^{10}$Center for Particle and Gravitational Astrophysics,
Institute for Gravitation and the Cosmos, Pennsylvania State University,
525 Davey Laboratory, University Park, PA 16802, USA\\
$^{11}$Physics Department, Indiana University, Bloomington, IN 47405, USA; kostelec@indiana.edu}

\begin{abstract}
Violations of Lorentz invariance
can lead to an energy-dependent vacuum dispersion of light,
which results in arrival-time differences of photons
arising with different energies from a given transient source.
In this work,
direction-dependent dispersion constraints are obtained
on nonbirefringent Lorentz-violating effects,
using the observed spectral lags of the gamma-ray burst GRB 160625B.
This burst has unusually large high-energy photon statistics,
so we can obtain constraints
from the true spectral time lags of bunches of high-energy photons
rather than from the rough time lag of a single highest-energy photon.
Also,
GRB 160625B is the only burst to date having a well-defined transition
from positive lags to negative lags,
which provides a unique opportunity
to distinguish Lorentz-violating effects
from any source-intrinsic time lag in the emission of photons
of different energy bands.
Our results place comparatively robust two-sided constraints
on a variety of isotropic and anisotropic coefficients for Lorentz violation,
including first bounds on Lorentz-violating effects
from operators of mass dimension ten in the photon sector.
\end{abstract}


\section{Introduction}
\label{sec:intro}

Lorentz invariance is the foundational symmetry
of Einstein's relativity.
However,
deviations from Lorentz symmetry
at the Planck energy scale
$E_{\rm Pl}=\sqrt{\hbar c^{5}/G}\simeq1.22\times10^{19}$ GeV
are predicted in various quantum gravity theories
attempting to unify General Relativity and quantum mechanics
\citep{1989PhRvD..39..683K,
1991NuPhB.359..545K,
1995kp,
2005LRR.....8....5M,
2006rtb,
2013LRR....16....5A,
2014jdt}.
The prospect of discovering Lorentz violation in nature
via sensitive relativity tests
has motivated numerous recent experimental searches.
A compilation of results can be found in
\cite{2011RvMP...83...11K}.

Although any deviations from Lorentz symmetry
are expected to be tiny at attainable energies $\ll E_{\rm Pl}$,
they can become detectable when particles travel over large distances.
Astrophysical observations involving long baselines can therefore provide
exceptionally sensitive tests of Lorentz invariance.
In the photon sector,
signatures of Lorentz violation
include vacuum dispersion and vacuum birefringence,
along with direction-dependent effects
\citep{2008ApJ...689L...1K}.
Vacuum dispersion produces a frequency-dependent velocity of the photon.
Lorentz invariance can therefore be tested
by comparing the arrival time differences of photons at different wavelengths
originating from the same astrophysical source
(see, e.g.,
\citealt{1998Natur.393..763A,
2005PhLB..625...13P,
2006APh....25..402E,
2008JCAP...01..031J,
2008ApJ...689L...1K,
2009PhRvD..80a5020K,
2009Sci...323.1688A,
2013PhRvD..87l2001V,
2013APh....43...50E,
2015PhRvD..92d5016K,
2017ApJ...834L..13W}).
Similarly,
vacuum birefringence results in an energy-dependent rotation
of the polarization plane of linearly polarized photons.
Thus,
astrophysical polarization measurements
can also be used to test Lorentz invariance
(see, e.g.,
\citealt{
2001km,
2006km,
2007PhRvL..99a1601K,
2013PhRvL.110t1601K,
2009JCAP...08..021G,
2011APh....35...95S,
2011PhRvD..83l1301L,
2012PhRvL.109x1104T,
2017PhRvD..95h3013K}).
Since polarization measurements are more sensitive
than vacuum dispersion time-of-flight measurements
by a factor $\propto 1/E$,
where $E$ is the energy of the light,
polarization measurements typically yield more stringent limits
(\citealt{2009PhRvD..80a5020K}).
However,
many predicted signals of Lorentz violation
have no vacuum birefringence,
so limits from time-of-flight measurements are essential
in a broad-based search for effects.

At attainable energies,
violations of Lorentz invariance are described
by the Standard-Model Extension (SME),
which is the comprehensive realistic effective field theory
characterizing Lorentz and CPT violation
\citep{1997PhRvD..55.6760C,
1998PhRvD..58k6002C,
2004PhRvD..69j5009K}.
Each term in the SME Lagrange density
consists of a Lorentz-violating operator of definite mass dimension $d$
in natural units ($\hbar = c = 1$),
contracted with a coefficient that governs the size
of any observable effects.
For arbitrary $d$,
all terms affecting photon propagation
have been constructed explicitly
\citep{2009PhRvD..80a5020K}.
Photon vacuum dispersion is induced by operators of dimension $d\neq 4$
and is proportional to $(E/E_{\rm Pl})^{d-4}$.
Lorentz-violating terms with odd $d$ violate CPT symmetry,
while those with even $d$ preserve CPT.
For odd $d$,
vacuum dispersion and birefringence always occur together,
whereas for each even $d$
there is a subset of $(d-1)^{2}$
nonbirefringent but dispersive Lorentz-violating operators.
The latter form an ideal target for time-of-flight measurements.
In this paper,
we focus on measuring coefficients controlling nonbirefringent dispersion
with even $d=6$, $8$, and 10.

In a previous paper
\citep{2017ApJ...834L..13W},
we derived new limits on isotropic linear and quadratic
leading-order Lorentz-violating vacuum dispersion
using the gamma-ray burst GRB 160625B,
which is the only one to date
known to display a well-defined transition
from positive spectral lags to negative spectral lags.
Spectral lag is the arrival-time difference
either of a given feature such as a peak
in light curves from the same source in different energy bands
or of high- and low-energy photons,
and it is a common observational feature in gamma-ray bursts
(see, e.g.,
\citealt{1995A&A...300..746C,
1996ApJ...459..393N,
1997ApJ...486..928B}).
In our conventions,
a positive spectral lag corresponds to
an earlier arrival time for the higher-energy photons.
The restriction to isotropic vacuum dispersion
disregards $d(d-2)$ possible effects
from anisotropic violations at each $d$ in a nonbirefringent scenario.
In this paper,
we use the peculiar time-of-flight measurements of GRB 160625B
to constrain combinations of nonbirefringent Lorentz-violating coefficients
with mass dimension $d=6$, $8$, and $10$,
allowing for all direction-dependent effects.

To date,
limits on the 25 $d=6$ nonbirefringent coefficients for Lorentz violation
have been obtained by studying the dispersion of light
in observations of GRB 021206
\citep{2004ApJ...611L..77B,
2008ApJ...689L...1K},
GRB 080916C
\citep{2009Sci...323.1688A,
2009PhRvD..80a5020K},
GRB 090510
\citep{2010arXiv1008.2913F},
four bright gamma-ray bursts
\citep{2013PhRvD..87l2001V},
the blazar Markarian 501
\citep{2008PhLB..668..253M,
2008ApJ...689L...1K},
the active galaxy PKS 2155-304
\citep{2008PhRvL.101q0402A,
2009PhRvD..80a5020K},
and 25 active galaxy nuclei
\citep{2015PhRvD..92d5016K}.
Only three bounds have been obtained
on combinations of the 49 $d=8$ coefficients
for nonbirefringent vacuum dispersion,
derived from GRB 021206
\citep{2004ApJ...611L..77B,
2008ApJ...689L...1K},
GRB 080916C
\citep{2009Sci...323.1688A,
2009PhRvD..80a5020K},
and GRB 090510
\citep{2010arXiv1008.2913F}.
No constraints have been placed on the 81 $d=10$ coefficients.
A compilation of the current limits in the literature
can be found in \cite{2011RvMP...83...11K}.
While these constraints have reached high precision,
most are obtained by concentrating on the time delay induced
by Lorentz violation
and neglecting the intrinsic time delays
that depend on the emission mechanism of the astrophysical sources.
Furthermore,
the limits from gamma-ray bursts are based on the rough time lag
for a single GeV-scale photon.
Performing a search for Lorentz violation
using true time lags of high-quality and high-energy light curves
in multi-photon bands of different energy
is therefore both timely and crucial.

In this work,
by fitting the true multi-photon spectral-lag data of GRB 160625B,
we give both
a plausible description of the intrinsic energy-dependent time lag
and robust constraints on Lorentz-violating coefficients
with $d=6$, $8$, and (for the first time) $10$.
In Section~\ref{sec:formalism},
we present an overview of the theoretical foundation of vacuum dispersion
due to Lorentz violation.
The data analysis and our results
constraining coefficients for Lorentz violation
are presented in Section~\ref{sec:LIV}.
A brief summary and discussion are provided in Section~\ref{sec:summary}.

\section{Vacuum Dispersion in the Standard-Model Extension}
\label{sec:formalism}

In the SME framework,
the modified dispersion relations
for photon propagation {\it in vacuo} take the form
\citep{2008ApJ...689L...1K,
2009PhRvD..80a5020K}
\begin{equation}
  E(p) \simeq
\bigl(1 - \varsigma^0
\pm \sqrt{(\varsigma^1)^2 + (\varsigma^2)^2 + (\varsigma^3)^2}\bigr)
\, p\;,
\label{mdr}
\end{equation}
where $p$ is the photon momentum.
The quantities $\varsigma^0$, $\varsigma^1$, $\varsigma^2$, and $\varsigma^3$
are momentum- and direction-dependent combinations
of coefficients for Lorentz violation
that can be decomposed in a spherical basis to yield
\begin{equation}
\begin{aligned}
  \varsigma^0 &=
\sum_{djm}p^{d-4} {}_{0}Y_{jm}(\hat{\textbf{\emph{n}}})c_{(I)jm}^{(d)},
\\
  \varsigma^1 \pm i\varsigma^2 &=
\sum_{djm}p^{d-4} {}_{\mp2}{Y}_{jm}(\hat{\textbf{\emph{n}}})
\left(k_{(E)jm}^{(d)} \mp ik_{(B)jm}^{(d)}\right),
\\
  \varsigma^3 &=
\sum_{djm}p^{d-4} {}_{0}Y_{jm}(\hat{\textbf{\emph{n}}})k_{(V)jm}^{(d)},
\label{eq:sigma0}
\end{aligned}
\end{equation}
where $\hat{\textbf{\emph{n}}}$ points towards the source
and ${}_{s}{Y}_{jm}(\hat{\textbf{\emph{n}}})$
are spin-weighted harmonics of spin weight $s$.
The standard spherical polar coordinates $(\theta, \phi)$
associated with $\hat{\textbf{\emph{n}}}$
are defined in a Sun-centered celestial-equatorial frame
\citep{2002PhRvD..66a6005K},
with $\theta = (90^{\circ}-\rm{Dec.})$ and $\phi =$ R.A.,
where the astrophysical source
is at right ascension R.A.\ and declination Dec.\

The above decomposition characterizes
all types of Lorentz violations for vacuum propagation
in terms of four sets of spherical coefficients.
For even $d$,
the coefficients are
$c_{(I)jm}^{(d)}$,
$k_{(E)jm}^{(d)}$,
$k_{(B)jm}^{(d)}$
and control CPT-even effects.
For odd $d$,
the coefficients are
$k_{(V)jm}^{(d)}$
and govern CPT-odd effects.
For example,
in the isotropic limit the group-velocity defect for photons
is found to be
\begin{equation}
\delta v_g \simeq
\frac{1}{\sqrt{4\pi}} \sum_d (d-3) E^{d-4}
\big(- c_{(I)00}^{(d)} \pm k_{(V)00}^{(d)} \big)\;,
\end{equation}
where the factor $(d-3)$ reflects the difference between group
and phase velocities and is included here because its
significance grows with larger values of $d$.
Birefringence results when the usual degeneracy among polarizations is broken,
for which at least one of
$k_{(E)jm}^{(d)}$,
$k_{(B)jm}^{(d)}$,
or $k_{(V)jm}^{(d)}$
is nonzero.
The only coefficients for nonbirefringent vacuum dispersion
are $c_{(I)jm}^{(d)}$,
which are therefore the focus of the present work.
Note that all the spherical coefficients
can be taken to be constants in the Sun-centered frame.

For even $d>4$,
nonzero values of $c_{(I)jm}^{(d)}$
lead to an energy dependence of the photon velocity in vacuum,
so two photons of different energies $E_{\rm h}>E_{\rm l}$ emitted simultaneously
from an astrophysical source at redshift $z$ would arrive at Earth
at different times.
Setting to zero the coefficients for birefringent propagation,
the group-velocity defect is given by
\begin{equation}
\delta v_g = - \sum_{djm} (d-3)
E^{d-4} \, {}_{0}Y_{jm}(\hat{\textbf{\emph{n}}}) c_{(I)jm}^{(d)}\;,
\end{equation}
which includes direction-dependent effects.
The induced arrival-time difference
can therefore be written as
\begin{equation}
\begin{aligned}
\Delta t_{\rm LV}&=t_{\rm l}-t_{\rm h}
\\
&
\hskip -20pt
\approx
-(d-3)\left(E_{\rm h}^{d-4} - E_{\rm l}^{d-4}\right)
\int_0^{z}\frac{(1 + z')^{d-4}}{H_{z'}} {\rm d} z'
\sum_{jm} {}_{0}Y_{jm}(\hat{\textbf{\emph{n}}}) c_{(I)jm}^{(d)}
\;,
\label{eq:tLIV}
\end{aligned}
\end{equation}
where $t_{\rm h}$ and $t_{\rm l}$ are the arrival times
of the high-energy photons and the low-energy photons,
respectively.
Also,
$H_z = H_0\left[\Omega_m(1+z)^3 + \Omega_\Lambda\right]^{1/2}$
is the Hubble expansion rate at $z$,
where the standard flat $\Lambda$CDM model
with parameters
$H_{0}=67.3$ km $\rm s^{-1}$ $\rm Mpc^{-1}$,
$\Omega_{m}=0.315$,
and $\Omega_{\Lambda}=1-\Omega_{m}$
is adopted
\citep{2014A&A...571A..16P}.
Note that the coefficients $c_{(I)jm}^{(d)}$
can be either positive or negative,
contributing to a decrease or an increase in photon velocity
with increasing photon energy,
respectively.
For example,
when
$\sum_{jm} {}_{0}Y_{jm}(\hat{\textbf{\emph{n}}}) c_{(I)jm}^{(d)}$
is positive,
photons with higher energies would arrive on Earth
after those with lower ones,
implying a negative spectral lag due to Lorentz violation.

\section{Constraints on SME Coefficients}
\label{sec:LIV}

\begin{figure}
\vskip-0.1in
\centerline{\includegraphics[angle=0,width=1.15\hsize]{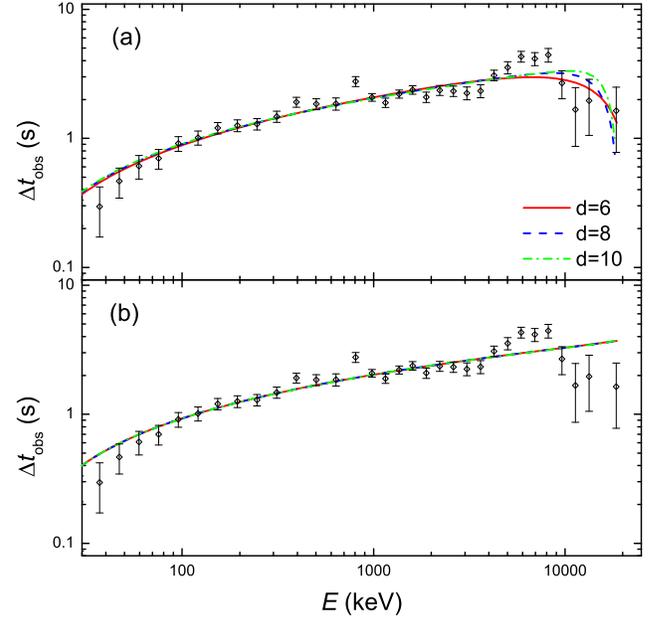}}
\vskip-0.1in
\caption{
Panel (a):
energy dependence of the observed spectral lag $\Delta t_{\rm obs}$
of the second sub-burst of GRB 160625B
relative to the lowest-energy band,
and the best-fit theoretical curves for the negative-lag case.
Solid line: model with $d=6$ coefficients.
Dashed line: model with $d=8$ coefficients.
Dash-dotted line: model with $d=10$ coefficients.
Panel (b):
same but for the positive-lag case.}
\label{f1}
\end{figure}

In this section,
we use the observation of GRB 160625B
to place direction-dependent bounds
on combinations of the coefficients
$c_{(I)jm}^{(6)}$,
$c_{(I)jm}^{(8)}$,
and $c_{(I)jm}^{(10)}$.
GRB 160625B was detected by the \emph{Fermi} satellite
on 2016 June 25 at $T_{0}=22:40:16.28$ UT,
with coordinates (J2000) R.A.=$308^{\circ}$ and Dec.=$+6.9^{\circ}$
\citep{2016GCN..19581...1B,
2016GCN..19580...1D}.
Its redshift is $z=1.41$
\citep{2016GCN..19600...1X}.
This gamma-ray burst is special because its gamma-ray light curve
consists of three dramatically different isolated sub-bursts
with very high photon statistics
\citep{2016GCN..19581...1B,
2016arXiv161203089Z}.
Since the second sub-burst of GRB 160625B is extremely bright,
its light curve in different energy bands can be readily extracted.
With multi-photon energy bands,
\cite{2017ApJ...834L..13W}
calculated the spectral time lags in the light curves
recorded in the lowest-energy band (10--12 keV)
relative to any of the other light curves in higher-energy bands,
finding that the observed lag $\Delta t_{\rm obs}$
increases at $E\la8$ MeV
and then gradually decreases in the energy range
8 MeV $\la  E\la$ 20 MeV.
Table~1 of \cite{2017ApJ...834L..13W} contains
the 37 energy-lag measurements obtained from this analysis.
The lag behavior is shown in Fig.~\ref{f1}
and is very peculiar,
being the first transition from positive to negative lags
discovered within a burst.
Note that the spectral-lag analysis is restricted
to the data from the Gamma-Ray Monitor on board the $Fermi$ satellite.
In principle,
stronger constraints on coefficients for Lorentz violation
could be derived by including the higher-energy photons
detected by the $Fermi$ Large Area Telescope (LAT).
However,
the LAT data are much sparser in terms of photon number,
and there are only ten $>1$ GeV LAT photons
during the second sub-burst of GRB 160625B
\citep{2016arXiv161203089Z}.
This makes it challenging to extract the LAT-band light curve
with high temporal resolution
and requires additional assumptions in the error analysis
beyond those adopted here.
A future large set of higher-energy data
could permit further improvements over the results we report below.

Since the time delay
$\Delta t_{\rm LV}$
induced by Lorentz violation
is likely to be accompanied
by an intrinsic energy-dependent time delay
$\Delta t_{\rm int}$
caused by unknown properties of the source
(see, e.g.,
\citealt{2006APh....25..402E,
2009CQGra..26l5007B}),
the observed time lag between two different energy bands
should include two parts,
\begin{equation}
\Delta t_{\rm obs}=\Delta t_{\rm int} + \Delta t_{\rm LV} \;.
\label{eq:tobs}
\end{equation}
As the spectral lags of most gamma-ray bursts
have a positive energy dependence,
with high-energy photons arriving earlier than the low-energy ones
(see
\citealt{2014AstL...40..235M,
2016arXiv161007191S}),
we propose that the observer-frame relationship
between the intrinsic time lag and the energy $E$
is approximately a power law with positive dependence,\footnote{
Statistically, the intrinsic time lags of most GRBs increase with
the energies E in the form of an approximate power-law function
(e.g., \citealt{2006APh....25..402E,2009CQGra..26l5007B}), i.e., the power-law
model is in fact an accurate representation of the energy dependence
of the spectral lag.}
\begin{equation}
\Delta t_{\rm int}(E)=
\tau
\left[
\left(\frac{E}{\rm keV}\right)^{\alpha}
-\left(\frac{E_{\rm l}}{\rm keV}\right)^{\alpha}
\right]\;{\rm s} \;,
\label{eq:tint}
\end{equation}
where $\tau>0$ and $\alpha>0$,
and where $E_{\rm l}=11.34$ keV is the median value
of the fixed lowest-energy band (10--12 keV).
Also,
Lorentz violation with positive
$\sum_{jm} {}_{0}Y_{jm}(\hat{\textbf{\emph{n}}}) c_{(I)jm}^{(d)}$
implies high-energy photons arrive later than low-energy ones,
so the positive correlation between the lag and the energy
should gradually turn negative
as the Lorentz violation becomes dominant at higher energies.
The combined contributions from the intrinsic time lag
and the Lorentz-violation lag
can therefore lead to the observed lag behavior
with a transition from positive to negative lags
\citep{2017ApJ...834L..13W}.

For the present analysis,
we fit Eqs.~(\ref{eq:tLIV}), (\ref{eq:tobs}), and (\ref{eq:tint})
to the 37 energy-lag measurements of GRB 160625B
using the standard minimum $\chi^{2}$ statistic.
The parameters $\tau$ and $\alpha$ are allowed to be free
and are fitted simultaneously with the combination
$\sum_{jm} {}_{0}Y_{jm}(\hat{\textbf{\emph{n}}}) c_{(I)jm}^{(d)}$
of SME coefficients
to yield best-fit values and uncertainties.
Since the scenarios with positive and negative spectral lag
due to Lorentz violation
produce qualitatively different curves,
we study the two cases separately.
The data from GRB 160625B
contain an apparent transition to a negative lag at high energies
(see Fig.~\ref{f1}),
so the strongest constraints can be expected
in fitting for Lorentz violation with positive lag.

\begin{figure}
\centerline{\includegraphics[angle=0,width=0.5\textwidth]{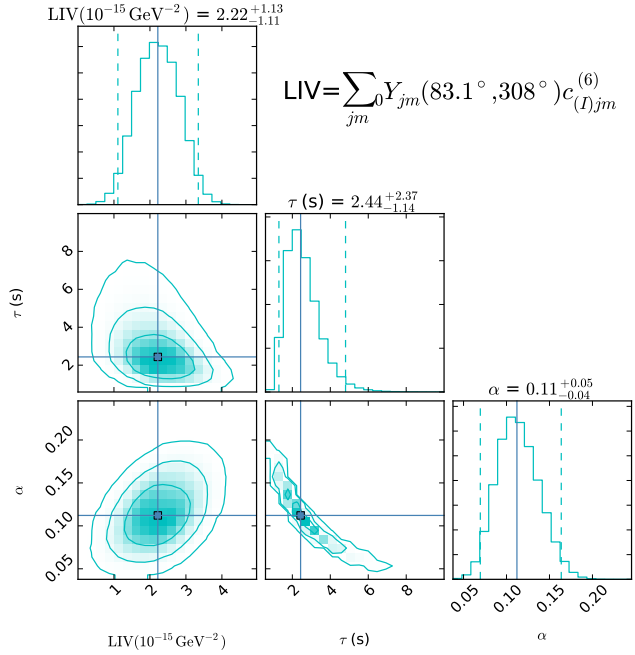}}
\vskip-0.1in
\caption{
1-D probability distributions and 2-D regions
with the $1\sigma$ to $3\sigma$ contours corresponding
to the parameters $\tau$, $\alpha$ and vacuum coefficients with $d=6$
for the negative-lag case.
The vertical solid lines indicate the best-fit values.
Made with triangle.py from \cite{2013PASP..125..306F}.}
\label{f2}
\vskip 0.1in
\end{figure}

\begin{figure}
\centerline{\includegraphics[angle=0,width=0.5\textwidth]{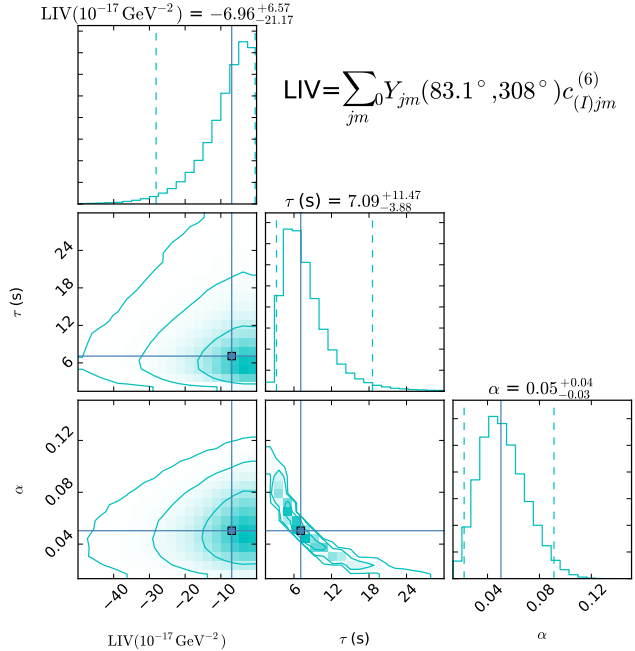}}
\vskip-0.1in
\caption{Same as Fig.~\ref{f2}, but for the positive-lag case.}
\label{f3}
\vskip 0.1in
\end{figure}

\begin{table*}
\caption{\rm Fit results and estimated 95\% C.L. constraints on coefficients
for $d=6$, 8, and 10.
\label{t1}}
\begin{center}
\vskip -12pt
\setlength{\tabcolsep}{4pt}
\begin{tabular}{ccccc}
\hline
\hline
&&&&\\[-8pt]																		
	&	$	d	$	&	$	6	$	&	$	8	$	&	$	10	$	\\	[3pt]
Negative spectral lag	&	$	\tau	$	&	$	2.44^{+2.37}_{-1.14}	$	&	$	3.50^{+3.75}_{-1.69}	$	&	$	4.15^{+4.85}_{-2.07}	$	 \\	[3pt]
	&	$	\alpha	$	&	$	0.11^{+0.05}_{-0.04}	$	&	$	0.09^{+0.05}_{-0.04}	$	&	$	0.08^{+0.05}_{-0.03}	$	\\	[3pt]
	&	$	\sum_{jm}\phantom{}_{0}Y_{jm}(83.1{^\circ},308{^\circ})c^{(d)}_{(I)jm}	$	&	$	2.22^{+1.13}_{-1.11}\times10^{-15}\; \rm GeV^{-2}	$	 &	$	1.38^{+0.65}_{-0.65}\times10^{-12}\; \rm GeV^{-4}	$	&	$	 6.38^{+3.60}_{-3.60}\times10^{-10}\; \rm GeV^{-6}	$	\\	[3pt]
	&	$	\chi^2/{\rm d.o.f.}	$	&	$	76.70/34=2.26	$	&	$	74.41/34=2.19	$	&	$	77.95/34=2.29	$	\\	[6pt]
Positive spectral lag	&	$	\tau	$	&	$	7.09^{+11.47}_{-3.88}	$	&	$	6.94^{+10.92}_{-3.72}	$	&	$	6.92^{+11.00}_{-3.77}	$	 \\	[3pt]
	&	$	\alpha	$	&	$	0.05^{+0.04}_{-0.03}	$	&	$	0.05^{+0.04}_{-0.03}	$	&	$	0.05^{+0.04}_{-0.03}	$	\\	[3pt]
	&	$	\sum_{jm}\phantom{}_{0}Y_{jm}(83.1{^\circ},308{^\circ})c^{(d)}_{(I)jm}	$	&	$	-6.96^{+6.57}_{-21.17}\times10^{-17}\; \rm GeV^{-2}	$	 &	$	-3.71^{+3.50}_{-11.39}\times10^{-14}\; \rm GeV^{-4}	$	&	$	 -2.25^{+2.12}_{-6.81}\times10^{-11}\; \rm GeV^{-6}	$	\\	[3pt]
	&	$	\chi^2/{\rm d.o.f.}	$	&	$	93.25/34=2.74	$	&	$	93.25/34=2.74	$	&	$	93.25/34=2.74	$	\\	[3pt]
\hline&&&&\\[-8pt]																		
95\% C.L. bounds	&	$	-2.8\times10^{-16}\; {\rm GeV^{-2}}<	$	&	$	\sum_{jm}\phantom{}_{0}Y_{jm}(83.1{^\circ},308{^\circ})c^{(6)}_{(I)jm}	 $	&	$	<3.4\times10^{-15}\; {\rm GeV^{-2}}	$	&	$		$	\\	 [3pt]
	&	$	-1.0\times10^{-15}\; {\rm GeV^{-2}}<	$	&	$	c^{(6)}_{(I)00}	$	&	$	<1.2\times10^{-14}\; {\rm GeV^{-2}}	$	&	$		$	\\	 [3pt]
	&	$	-1.5\times10^{-13}\; {\rm GeV^{-4}}<	$	&	$	\sum_{jm}\phantom{}_{0}Y_{jm}(83.1{^\circ},308{^\circ})c^{(8)}_{(I)jm}	$	&	$	 <2.0\times10^{-12}\; {\rm GeV^{-4}}	$	&	$		$	\\	[3pt]
	&	$	-5.4\times10^{-13}\; {\rm GeV^{-4}}<	$	&	$	c^{(8)}_{(I)00}	$	&	$	<7.2\times10^{-12}\; {\rm GeV^{-4}}	$	&	$		$	\\	 [3pt]
	&	$	-9.1\times10^{-11}\; {\rm GeV^{-6}}<	$	&	$	\sum_{jm}\phantom{}_{0}Y_{jm}(83.1{^\circ},308{^\circ})c^{(10)}_{(I)jm}	$	&	$	 <1.0\times10^{-9}\; {\rm GeV^{-6}}	$	&	$		$	\\	[3pt]
	&	$	-3.2\times10^{-10}\; {\rm GeV^{-6}}<	$	&	$	c^{(10)}_{(I)00}	$	&	$	<3.5\times10^{-9}\; {\rm GeV^{-6}}	$	&	$		$	 \\	[3pt]
\hline
\hline
\end{tabular}
\end{center}
\end{table*}

The results of the various fits are presented in Table \ref{t1}.
The cases of negative and positive Lorentz-violating spectral lag
are displayed separately.
For each each value $d=6$, 8, 10 in turn,
the best-fit results and $2\sigma$ uncertainties
are provided for the parameters $\tau$ and $\alpha$
and for the combination
$\sum_{jm} {}_{0}Y_{jm}(\hat{\textbf{\emph{n}}}) c_{(I)jm}^{(d)}$
of coefficients for Lorentz violation,
along with the $\chi^2$ value for the fit.
The latter values imply that nonzero Lorentz violation
with $d=6$, 8, or 10 cannot be inferred from the data,
so the results are most usefully expressed as constraints.
The lower part of Table \ref{t1}
gives two-sided estimated constraints at the 95\% confidence level
on each of the direction-dependent combinations
of coefficients for Lorentz violation.
We also provide estimated constraints at the 95\% confidence level
for the limiting case of isotropic Lorentz violation,
where only the coefficients with $j=m=0$ are relevant.

To illustrate the fits,
the theoretical curves obtained from each of the best-fit values
are provided in Fig.~\ref{f1}
for Lorentz violation contributing to both negative and positive lag.
Also,
the probability distributions for the analysis with $d=6$ coefficients
are shown in Fig.~\ref{f2}
for the scenario of negative lag due to Lorentz violation
and in Fig.~\ref{f3} for positive lag.
These figures show
the one-dimensional probability distribution
for each free parameter
and the two-dimensional contour plots
with 1$\sigma$ to 3$\sigma$ confidence regions
for the two-parameter combinations.
The corresponding distributions
for the $d=8$ and 10 cases are qualitatively similar.
As expected,
the spreads of the distributions involving
negative values of the combination
$\sum_{jm} {}_{0}Y_{jm}(\hat{\textbf{\emph{n}}}) c_{(I)jm}^{(d)}$
are much larger than those for positive values,
reflecting the occurrence at high energies
of the negative lag transition in the data.

\section{Summary and Discussion}
\label{sec:summary}

In this work,
we use data from GRB 160625B
to constrain Lorentz violation in the photon sector.
The general description of possible photon behaviors,
allowing for operators of arbitrary mass dimension $d$,
is predicted in the SME effective field theory
for Lorentz violation.
The dispersion relation for photons
can acquire corrections that depend on polarization,
direction of propagation, and energy.
Here,
we study nonbirefringent effects
yielding energy- and direction-dependent vacuum dispersion
and controlled by coefficients for Lorentz violation with $d=6$, 8, and 10.

Gamma-ray bursts are well suited for these studies
because they are distant transient sources
involving a range of photon energies
and so permit time-of-flight tests.
A key challenge in this approach
is to distinguish an intrinsic time delay at the source
from a time delay induced by Lorentz violation.
Most previous studies limit attention to the latter
while ignoring possible source-intrinsic effects,
which would impact the reliability of the resulting constraints
on Lorentz violation.
Furthermore,
prior limits from gamma-ray bursts are based
on the approximate time delay of a single GeV-scale photon.
To obtain reliable constraints on Lorentz violation,
it is desirable to use the true time delays of broad light curves
in different energy multi-photon bands.

GRB 160625B has unusually high photon statistics,
allowing the use of amply populated energy bands.
Moreover,
it is the only burst so far with a well-defined
transition from positive to negative spectral lag.
This provides a unique opportunity
not only to disentangle the intrinsic time delay problem
but also to place more reliable constraints on Lorentz violation.
Here,
we propose that the intrinsic time delay
has a positive dependence on the photon energy.
Lorentz violation can cause high-energy photons
to travel slower than low-energy ones in the vacuum,
so the positive correlation of the time delay with energy
would gradually become an anticorrelation
as the Lorentz violation becomes dominant at higher energies.
By fitting the spectral lag behavior of GRB 160625B,
we obtain both a reasonable formulation
of the intrinsic energy-dependent time delay
and comparatively robust two-sided limits
on coefficients for Lorentz violation,
as shown in Table \ref{t1}.

Existing limits on coefficients for Lorentz violation,
including photon-sector constraints using other astrophysical sources
with the dispersion method,
are compiled in \cite{2011RvMP...83...11K}.
For the $d=6$ case,
our constraints are somewhat weaker or comparable to existing bounds
but can be viewed as comparatively robust.
For $d=8$,
only a few limits exist on the 49 coefficients
controlling nonbirefringent dispersion.
Our new bound is linearly independent of these
and so helps to constrain the full coefficient space.
For $d=10$,
our constraints are the first in the literature.
Note that
all bounds to date are based on photons with energies far below
the Planck scale, and so in principle there is still room for
Planck-scale effects.

The 95\% C.L.\ constraints given in Table \ref{t1}
assume no Lorentz violation in nature and
are extracted by considering only one value of $d$ at a time,
following the standard practice in the field.
However,
the presence of multiple values of $d$ could improve the fit.
As a proof of principle for this idea,
we have performed an analysis allowing
the $d=6$, 8, and 10 combinations of coefficients to vary simultaneously.
The best-fit values for the three coefficient combinations
are found to be approximately
$-8.97^{+4.38}_{-4.20}\times 10^{-15}$ GeV$^{-2}$,
$1.86^{+0.79}_{-0.77}\times 10^{-11}$ GeV$^{-4}$,
and $-7.00^{+3.23}_{-3.31}\times 10^{-9}$ GeV$^{-6}$,
respectively,
with $\chi^2=1.89$ per degree of freedom.
It is interesting to note that these nonzero best-fit values
are compatible with existing constraints in the literature.
Once other gamma-ray bursts displaying spectral-lag transitions
become available for similar studies,
a definitive search along these lines would become possible.
To discover further bursts with spectral-lag transitions,
we need to detect larger numbers of GRBs with higher temporal resolutions
and more high-energy photons.
The upcoming detectors for gamma-ray observations at very high energies
and with higher sensitivity and wider field-of-view,
such as the Large High Altitude Air Shower Observatory
\citep{2010ChPhC..34..249C,2014NIMPA.742...95C},
will be able to detect many high-energy ($>100$ GeV) photons
for each luminous GRB.
With large statistics for high-energy photons,
high-energy light curves with excellent temporal resolutions
can be constructed,
so the discovery of additional bursts with spectral-lag transitions
can be expected.
In any case,
the future is bright for improving constraints on Lorentz violation
using time-of-flight tests from gamma-ray bursts.

\acknowledgments
This work is supported by the National Basic Research Program
(``973'' Program) of China (Grant No.\ 2014CB845800),
the National Natural Science Foundation of China
(Grant Nos.\ 11673068, 11603076, and 11103083),
the Youth Innovation Promotion Association (2011231 and 2017366),
the Key Research Program of Frontier Sciences (QYZDB-SSW-SYS005),
the Strategic Priority Research Program
``Multi-waveband gravitational wave Universe''
(Grant No.\ XDB23000000) of the Chinese Academy of Sciences,
the Natural Science Foundation of Jiangsu Province (Grant No.\ BK20161096),
and the Guangxi Key Laboratory for Relativistic Astrophysics.
B.B.Z.\ acknowledges support from
the Spanish Ministry Projects AYA 2012-39727-C03-01 and AYA2015-71718-R.
Part of this work used B.B.Z.'s personal IDL code library ZBBIDL
and personal Python library ZBBPY.
The computation resources used in this work
are owned by Scientist Support LLC.
L.S.\ acknowledges support from
the Joint NSFC-ISF Research Program (No.\ 11361140349),
jointly funded by the National Natural Science Foundation of China
and the Israel Science Foundation.
P.M.\ acknowledges NASA NNX 13AH50G.
V.A.K.\ acknowledges support from the United States Department of Energy
under grant {DE}-SC0010120
and from the Indiana University Center for Spacetime Symmetries.

\bibliographystyle{apj}

\end{document}